\RequirePackage{fix-cm}
\documentclass{svjour3}                     
\smartqed  
\usepackage[top=2.5cm,right=3cm,bottom=2.5cm,left=2.5cm]{geometry}
\usepackage{graphicx}
\usepackage{caption}
\usepackage{array,multirow}
\usepackage{amssymb, amsmath}
\usepackage{dutchcal}
\usepackage{braket}
\usepackage{calrsfs}
\DeclareMathAlphabet{\pazocal}{OMS}{zplm}{m}{n}
\usepackage{xfrac}
\usepackage{mathptmx}      
\journalname{Few Body Systems}
\begin{document}

\title{Determination of parameters of a potential model for tetraquark study by studying all S-wave mesons}

\author{Zheng Zhao \and Kai Xu\and Attaphon Kaewsnod\and Xuyang Liu\and Ayut Limphirat\and Yupeng Yan
}

\authorrunning{Zheng Zhao \and Kai Xu\and Attaphon Kaewsnod\and Xuyang Liu\and Ayut Limphirat\and Yupeng Yan}

\institute{Z. Zhao \and K.Xu \and A. Kaewsnod\and A. Limphirat\and Y. Yan \at
              School of Physics and Center of Excellence in High Energy Physics and Astrophysics, Suranaree University of Technology, Nakhon Ratchasima 30000, Thailand\\
              \email{zhaozheng1022@hotmail.com, yupeng@g.sut.ac.th}
               \and
               X. Liu \at
               School of Physics, Liaoning University, Shenyang 110036, China\\
               }

\date{Received: date / Accepted: date}

\maketitle

\begin{abstract}
The masses of low-lying S-wave mesons are evaluated in a constituent quark model (CQM) where the Cornell-like potential and one-gluon exchange spin-spin interaction are employed. To make the model applicable to both the light and heavy quark sectors, we introduce mass-dependent coupling coefficients. There are four free parameters in the model, which are determined by comparing the theoretical results with experimental data. The established model with one set of parameters may be applied to study higher excited meson states as well as multiquark systems in both the light and heavy quark sectors.

\end{abstract}

\section{Introduction}
\label{intro}

\indent Charged charmonium-like particles like $Z^+_c(3900)$, $Z^+_c(4020)$, $Z^+_c(4050)$, $Z^+_c(4055)$, $Z^+_c(4100)$, $Z^+_c(4200)$, $Z^+_c(4250)$, and $Z^+_c(4430)$ have been successively observed by experimental collaborations \cite{Brambilla:2019esw}. These charged charmonium-like states are beyond the conventional $c\bar c$-meson picture. Because of carrying one charge, these states are likely tetraquark systems with a quark content $c\bar cu\bar d$. Recently, a new resonance named $Z_{cs}(3985)^-$ has been observed by BESIII Collaboration \cite{Ablikim:2020hsk}, which is the first candidate of the charged charmonium-like tetraquark with strangeness, with a quark content $c\bar cs\bar u$. The significance of the resonance hypothesis is estimated to be 5.3$\sigma$. Later, the LHCb Collaboration has reported four exotic states $Z_{cs}(4000)^+$, $Z_{cs}(4220)^+$, $X(4685)$, and $X(4630)$ with a quark content $c\bar cu\bar s$ decaying to the $J/\psi K^+$ final state with high significance \cite{Aaij:2021ivw}. The first all-heavy multiquark exotic candidates $X(6900)$ with a quark content $c\bar cc\bar c$ has been recently observed by LHCb Collaboration in the $J/\psi$-pair mass spectrum \cite{Aaij:2020fnh}.

\indent In addition to the large number of XYZ tetraquark candidates, the observations of four pentaquark-like states has been reported by LHCb Collaboration. $P_c(4380)^+$ and $P_c(4450)^+$ have been observed for the first time in the process $\Lambda^0_b\to J/\psi pK^-$, with more than 9$\sigma$ significance  \cite{Aaij:2015tga,Aaij:2016phn}.
In recent years, another narrow pentaquark state $P_c(4312)^+$ is observed in the process $\Lambda^0_b\to J/\psi pK^-$ with a statistical significance is 7.3$\sigma$, and the $P_c(4450)^+$ pentaquark structure previously reported by LHCb has been confirmed and resolved at 5.4$\sigma$ significance into two narrow states: the $P_c(4440)^+$ and the $P_c(4457)^+$ \cite{Aaij:2019vzc}. All four pentaquark-like states may have the quark content of $uudc\bar c$.

\indent Various phenomenological research methods of hadron physics can be tested due to these observed multiquark states. A systematic spectrum study of the multiquark states would provide information for future experimental search for the missing higher excitations.

\indent The mass spectrum of charmonium-like tetraquark has been studied in non-relativistic potential model \cite{SilvestreBrac:1993ss,Patel:2014vua}, chromomagnetic interaction model \cite{Zhao:2014qva,Wu:2018xdi}, color flux-tube model \cite{Deng:2015lca}, relativized diquark model \cite{Anwar:2018sol}, and QCD sum rules \cite{Wang:2019mta}. Fully-heavy tetraquark mass spectrum has also been studied in non-relativistic quark model \cite{SilvestreBrac:1993ss,Brink:1998as,Berezhnoy:2011xn,Wang:2019rdo,Liu:2019zuc,Debastiani:2017msn,Yang:2020rih}, non-relativistic effective field theory \cite{Anwar:2017toa}, chromomagnetic interaction model \cite{Wu:2016vtq}, and QCD sum rules \cite{Wang:2018poa,Wang:2017jtz}. The pentaquark structure has been studied in various suggestions in the last decades. For the new observed $P_c$ states, there are three possible explanations for the structure: compact multiquarks \cite{Maiani:2015vwa,Lebed:2015tna,Wang:2015epa,Li:2015gta,Takeuchi:2016ejt}, hadronic molecules\cite{Roca:2015dva,He:2015cea,Chen:2015loa,Liu:2019tjn}
, and their admixtures\cite{Yamaguchi:2017zmn}.

\indent We are not to try to repeat the masses of the meson states by applying a large number of parameters but to develop a simple model with as less parameters as possible. We predetermine all model parameters by studying conventional $q\bar q$ mesons. The paper is organized as follows. The constituent quark model applied in our previous work \cite{Xu:2019fjt} is briefly reviewed in Sec.~\ref{GTM}. In Sec.~\ref{Re}, meson mass spectra are evaluated in the constituent quark model, and all model parameters are determined by comparing the theoretical and experimental masses of light, strange, charmed, and bottom mesons.

\section{Theoretical model}
\label{GTM}

\indent We study the multiquark systems in the nonrelativistic Hamiltonian,
\begin{flalign}\label{eqn::ham}
&H =H_0+ H_{hyp}^{OGE}, \nonumber \\
&H_{0} =\sum_{k=1}^{N} (\frac12M^{ave}_{k}+\frac{p_k^2}{2m_{k}})+\sum_{i<j}^{N}(-\frac{3}{16}\lambda^{C}_{i}\cdot\lambda^{C}_{j})(A_{ij} r_{ij}-\frac{B_{ij}}{r_{ij}}),  \nonumber \\
&H_{hyp}^{OGE} = -C_{OGE}\sum_{i<j}{\vec\lambda^{C}_{i}\cdot\vec\lambda^{C}_{j}}\,\vec\sigma_{i}\cdot\vec\sigma_{j},
\end{flalign}
by solving the $\rm Schr\ddot{o}dinger$ equation
\begin{flalign}\label{eqn::se}
H|\psi \rangle =E|\psi \rangle,
\end{flalign}
where the meson wave function $|\psi \rangle$ is expanded in the complete bases defined in Eq. (\ref{eqn::se}).
Here $m_k$ are the constituent quark masses taken from the previous work \cite{Xu:2020ppr},
\begin{eqnarray}\label{eq:nmo2}
m_u = m_d = 327 \ {\rm MeV}\,, \quad
m_s = 498 \ {\rm MeV}\,, \nonumber\\
m_c = 1642 \ {\rm MeV}\,, \quad
m_b = 4960 \ {\rm MeV}.\
\end{eqnarray}
$M^{ave}_k$ denotes the spin-averaged mass as $\frac{1}{4}M_{PS}+\frac{3}{4}M_V$ (except for $B_c$, $s\bar s$, $K^*$, and $q\bar q$), with $M_{PS}$ and $M_V$ being the mass of ground state pseudoscalar and vector mesons from experimental data \cite{PDG}. The spin-averaged masses $M_k^{ave}$ for each kind of mesons are listed in Table~\ref{tab:Maverage}.
\begin{table}
\centering
\caption{Spin-averaged masses $M_k^{ave}$ for various kinds of mesons. $M_{PS}$ and $M_V$ taken from PDG~\cite{PDG}.}
\label{tab:Maverage}
\begin{tabular}{lcccccccccc}
\hline\noalign{\smallskip}
Meson & $ c\bar c$ & $b\bar b $ & $B_c$ & $B_s$ & $B$ & $D_s$ & $D$ & $s\bar s$ & $K^*$ & $q\bar q$
\\
\noalign{\smallskip}\hline\noalign{\smallskip}
$[MeV]$&3068&9444&6323&5404&5314&2076&1972&955&819&683
\\
\noalign{\smallskip}\hline
\end{tabular}
\end{table}
$A_{ij}$ and $B_{ij}$ are mass-dependent coupling parameters, taking the form
\begin{eqnarray}
\label{aijbij}
A_{ij}= A_0 \sqrt{{m_{ij}}},\;\;B_{ij}=B_0 \sqrt{\frac{1}{m_{ij}}}.
\end{eqnarray}
with $m_{ij}$ being the reduced mass of $i$th and $j$th quarks, defined as $\;m_{ij}=\frac{2 m_i m_j}{m_i+m_j}$. The hyperfine interaction, $H_{hyp}^{OGE}$ includes only one-gluon exchange contribution \cite{Jaffe:1976ig}\cite{Jaffe:1976ih}. We follow the CQM convention and adopt $C_{OGE}=C_{ij}\frac{{m_u}^2}{m_im_j}$ \cite{Zhao:2014qva} with
\begin{eqnarray}
\label{cij}
 C_{ij} = C_{0}\left(\frac{m_{ij}}{m_c}\right)^{t}
\end{eqnarray}
where $C_{0}$ and $t$ are constants. $m_c$ is the constituent quark mass of c quark listed in Eq. (\ref{eq:nmo2}). $\vec\lambda^C_{i}$ and $\vec\sigma_i$ in Eq. (\ref{eqn::ham}) are the quark color operator and the spin operator respectively. In a meson, the contribution of the color part $\vec\lambda^C_{i}\cdot\vec\lambda^C_{j}$ in Eq. (\ref{eqn::ham}) is $-16/3$, and the contribution of $\vec\sigma_{i}\cdot\vec\sigma_{j}$ in Eq. (\ref{eqn::ham}) is $-3$ for $S=0$ and $+1$ for $S=1$ mesons. According to the hyperfine interaction form, we have
\begin{eqnarray}
\label{eqn::deltaM}
 \Delta M_{V-PS}=M_V-M_{PS}=C_{0}\left(\frac{m_{ij}}{m_c}\right)^{t} \frac{{m_u}^2}{m_im_j}(16+\frac{16}3)
\end{eqnarray}
with $m_{i}$ and $m_{j}$ being the reduced masses of $i$th and $j$th quark respectively. The constants $C_{0}$ and $t$ in Eq. (\ref{eqn::deltaM}) are fixed by applying the least squares method to minimize the weighted squared distance $\delta^2$,
\begin{eqnarray}
 \delta^2=\sum_{i=1}^{N}\omega_i\frac{(\Delta M_{V-PS}^{exp}-\Delta M_{V-PS}^{cal})^2}{{\Delta M_{V-PS}^{exp}}^2}
 \end{eqnarray}
where $\omega_i$ are weights being 1 for all the states. The $\Delta M_{V-PS}^{exp}$ is the experimental data with the form $\Delta M_{V-PS}^{exp}=M_V-M_{PS}$, and the $\Delta M_{V-PS}^{cal}$ is the theoretical results with the form $ \Delta M_{V-PS}^{cal}=C_{0}\left(\frac{m_{ij}}{m_c}\right)^{t} \frac{{m_u}^2}{m_im_j}(16+\frac{16}3)$. The fitting results of $\Delta M_{V-PS}$ are listed in Table~\ref{tab:hyp}.
\begin{table}
\centering
\caption{$\Delta M_{V-PS}$ for various kinds of mesons. $\Delta M_{V-PS}^{exp}$ taken from PDG~\cite{PDG}. Units are MeV.}
\label{tab:hyp}
\begin{tabular}{lcccccccccc}
\hline\noalign{\smallskip}
meson & $b\bar b $ & $ c\bar c$ & $B_s$ & $B$ & $D_s$ & $D$ \\
\noalign{\smallskip}\hline\noalign{\smallskip}
$\Delta M_{V-PS}^{exp}$ & 61 & 113 & 48 & 46 & 144 & 137
\\
$\Delta M_{V-PS}^{cal}$ & 52 & 113 & 57 & 52 & 138 & 135
\\
\noalign{\smallskip}\hline
\end{tabular}
\end{table}

\indent The spin-averaged mass of $B_c$ meson $M^{ave}_{Bc}$ takes the form $M_{PS}+16C_{ij}\frac{{m_u}^2}{m_im_j}$ because of the lack of $M_V$ experimental data. The spin-averaged masses of $s\bar s$, $K^*$ and $q\bar q$ take the form $M_{V}-(16/3)C_{ij}\frac{{m_u}^2}{m_im_j}$ to avoid the would-be Goldstone bosons of the chiral symmetry breaking.

\indent The four model coupling parameters in Eq. (\ref{aijbij}) and Eq. (\ref{cij}) are fixed by comparing theoretical results with experimental data \cite{PDG},
\begin{eqnarray}\label{eq:nmo1}
&A_0=3219.51 \ {\rm MeV^{3/2}}, \quad B_0=31.728 \ {\rm MeV^{1/2}}, \quad C_{0}=133.558 \ {\rm MeV}, \quad t=1.3
\end{eqnarray}
\indent In this work, we concentrate on the S-wave meson states and do not consider the tensor and spin-orbital interactions. The fitting results of meson states are shown in Sec.~\ref{Re}.

\section{Results and Summary}
\label{Re}
\indent We construct the complete bases by using the harmonic oscillator wave function. In our calculations, the bases size is $N=38$, and the length parameters of harmonic oscillator wave functions are adjusted to get the best eigenvalue of Eq. (\ref{eqn::se}).

\indent We calculate the mass spectra of light, strange, charmed, and bottom mesons which are believed mainly $q\bar q$ states in the Hamiltonian in Eq. (\ref{eqn::ham}). The theoretical results are collected, as shown in Table~\ref{tab:groundM}, with the experimental data taken from PDG~\cite{PDG} and some typical results of other works for comparison.

\indent In Ref \cite{SilvestreBrac:1993ss}, the S-wave tetraquark states with all quark configurations are systematically studied in a non-relativistic quark model. The parameters are fitted by reproducing all S-wave and P-wave ground state mesons. The charmonium-like tetraquark states $cu\bar c\bar d$ are systematically studied in a color flux-tube model with a multi-body confinement potential \cite{Deng:2015lca}. The parameters are determined by fitting the masses of the ground states of mesons. The mass spectrum of the ground state mesons is obtained by solving the two-body $\rm Schr\ddot{o}dinger$ equation. The fully-heavy tetraquark states $QQ\bar Q\bar Q$ are studied in several kinds of typical non-relativistic quark models in \cite{Wang:2019rdo,Liu:2019zuc,Debastiani:2017msn,Yang:2020rih}, where the Cornell-like potential are considered and model parameters are fixed by comparing the theoretical results of meson states with their experimental data.

\indent In summary, we are able to evaluate all ground and first radial excited meson states in a general model with reliable accuracy. Other works are applicable for either the heavy mesons or for the ground state mesons. By introducing three mass-dependent model parameters $A_{ij}$, $B_{ij}$, and $C_{ij}$ in Eq. (\ref{aijbij}) and Eq. (\ref{cij}), the number of parameters of our calculations is less than others, but the fitting results are compatible with both experimental data and other works. In future work, the Hamiltonian in Eq. (\ref{eqn::ham}) with the predetermined model parameters will be applied to predict the masses of multiquark states, tetraquarks and pentaquarks.

\begin{table}
\centering
\caption{Present the comparison of the theoretical results of low-lying meson states $M^{cal}$ with the experimental data $M^{exp}$ taken from PDG~\cite{PDG} and others. The D($\%$) column shows the deviation between the experimental and our theoretical mean values, $D=100\cdot (M^{exp}-M^{cal})/M^{exp}$. Units are MeV.}
\label{tab:groundM}
\begin{tabular}{lcccccccccccc}
\hline\noalign{\smallskip}
Meson & $M^{exp}$& $M^{cal}$  & D ($\%$) & \cite{SilvestreBrac:1993ss} & \cite{Deng:2015lca} & \cite{Wang:2019rdo} Model 1 & \cite{Wang:2019rdo} Model 2 & \cite{Liu:2019zuc} & \cite{Debastiani:2017msn} &\cite{Yang:2020rih}
\\
\noalign{\smallskip}\hline\noalign{\smallskip}
 $\Upsilon(1S)$ & 9460 & 9457 & 0.03 & 9433 & 9546 &  9503 & 9470 & 9460 & -- &9463
\\
 $\Upsilon(2S)$ & 10023 & 10046 &  -0.23 & -- & -- & 9949 & 10017 & 10017 & --& 9981
\\
 $\eta_b$ & 9399  & 9405 &  -0.06 & 9415 & 9441 & 9498 & 9428 & 9390 & --& 9401
\\
 $\eta_b(2S)$ & 9999 & 9994 &  0.05 & -- & -- & -- & -- & 9990 & --& 9961
 \\
\noalign{\smallskip}\hline\noalign{\smallskip}
 $J/\psi$ & 3097  &  3096 & 0.03 & 3097 &  3102 & 3085 & 3102 & -- & 3092 & 3102
\\
 $\psi(2S)$ & 3686 & 3686 &  0 & -- & -- & 3652 & 3658 & -- & 3671& 3720
\\
 $\eta_c$ & 2984 & 2983 &  0.03 & 3038 & 2912 & 3056 & 3006 & -- &2992 & 2968
\\
 $\eta_c(2S)$ & 3638  & 3573 &  1.79 & -- & -- & 3638 & 3621 & --&3632&3655
\\
 \noalign{\smallskip}\hline\noalign{\smallskip}
 $B_c$ & 6275 &  6275 &  0 & 6303 & 6261 &  6319 & 6293 & 6274 & -- & --
\\
 $B_c(2S)$ & 6842 & 6865 & -0.34 &-- & -- & -- &-- & 6842& -- & --
\\
\noalign{\smallskip}\hline\noalign{\smallskip}
 $B_s^*$ & 5415 & 5418 &  -0.06 & 5413 & 5430 &--&-- & --& -- & --
\\
 $B_s^0$ & 5367 & 5361 &  0.11 & 5372 & 5377 & --&-- & --& --& --
\\
\noalign{\smallskip}\hline\noalign{\smallskip}
 $B^*$ & 5325 & 5327 &  -0.04 & 5350 & 5301 & --&-- & --& --& --
\\
 $B^0$ & 5279 & 5275 &  0.08 & 5301 & 5259 & --&-- & --& --& --
\\
\noalign{\smallskip}\hline\noalign{\smallskip}
 $D_s^*$ & 2112 & 2110 &  0.09 & 2101 & 2140 & --&-- & --& --& --
\\
 $D_{s1}^*(2700)$ & 2708 & 2700 &  0.3 &-- & -- &--&-- & --& --& --
\\
 $D_s$ & 1968 & 1972 &  -0.2 & 1996 & 1972 & --&-- & --& --& --
\\
\noalign{\smallskip}\hline\noalign{\smallskip}
 $D^*(2007)^0$ & 2007 & 2006 &  0.05 & 2020 & 2002 & --&--& --& --& --
\\
 $D^0$ & 1870 & 1871 & -0.05 & 1886 & 1867 & --&--& --& --& --
\\
\noalign{\smallskip}\hline\noalign{\smallskip}
 $\phi(1020)$ & 1020 & 1019 &  0.1 & 1017 & 1112 & --&--& --& --& --
\\
 $\phi(1680)$ & 1680 & 1609 &  4.2 &-- & --&--&--& --& --& --
\\
\noalign{\smallskip}\hline\noalign{\smallskip}
 $K^*(892)$ & 892 & 892 &  0 & 905 & 974 & --&--& --& --& --
\\
 $K^*(1410)$ & 1414 & 1482 &  -4.8 &--&--&--&--& --& --& --
\\
\noalign{\smallskip}\hline\noalign{\smallskip}
 $\rho(770)$ & 770 & 770 &  0 & 777& 826&--&--& --& --& --
 \\
 $\rho(1450)$ & 1450 & 1359 &  6.3 & --&--&--&--& --& --& --
 \\
\noalign{\smallskip}\hline
\end{tabular}
\end{table}

\begin{acknowledgements}
This work is supported by Suranaree University of Technology (SUT). Z. Zhao acknowledges support from SUT-OROG Ph.D. scholarship under Contract No. 62/2559.  X.Y. Liu acknowledges support from the Young Science Foundation from the Education Department of Liaoning Province, China (Project No. LQ2019009).
\end{acknowledgements}



\end{document}